\documentclass{elsart}
\usepackage{graphicx}
\usepackage{epsfig}
\usepackage{relsize}  
\usepackage{bm}       
\usepackage{amsfonts} 
\usepackage{amsmath}  
\usepackage{amssymb}  
\usepackage{upgreek}  
\usepackage{xspace}   
\usepackage{array}
\usepackage[font=footnotesize,bf,labelsep=space]{caption}
\usepackage{slashed}  

\newcommand{\beq}{\begin{equation}}
\newcommand{\eeq}{\end{equation}}
\newcommand{\beqa}{\begin{eqnarray}}
\newcommand{\eeqa}{\end{eqnarray}}
\newcommand{\bseq}{\begin{subequations}}
\newcommand{\eseq}{\end{subequations}}

\newcommand{\trm}{\textrm}
\newcommand{\dd}{{\mathrm{d}}}
\newcommand{\tr}{{\mathrm{Tr }}}

\begin{document}
\begin{frontmatter}

\title{HTL approach to the viscosity of quark plasma}

\author{P. Czerski$^1$, W.M. Alberico$^{2,3}$,  S. Chiacchiera$^{2,3}$, 
A. De Pace$^3$, }
\author{H. Hansen$^{2,3,4}$, A. Molinari$^{2,3}$, M. Nardi$^3$}
\address{$^1${Institute of Nuclear Physics Polish Academy of Science,  Krak\'ow,\\
  ul. Radzikowskiego 152, Poland}\\
\ \\
  $^2$ Dipartimento di Fisica Teorica dell'Universit\`a di Torino \\ 
    via P.Giuria 1, I-10125 Torino, Italy\\
  \ \\
  $^3$ Istituto Nazionale di Fisica Nucleare, Sezione di Torino, \\ 
 via P.Giuria 1, I-10125 Torino, Italy\\
  \ \\
 $^4${{\tt Present address:} 
IPN Lyon, 43 Bd. du 11 Novembre 1918, F-69622, Villeurbanne, Cedex}}
\date{\today}

\begin{abstract}
The quark viscosity in the quark gluon plasma is evaluated in HTL approximation. The different 
contributions to the viscosity arising from the various components of the
quark spectral function are discussed. The calculation is extended to finite
values of the chemical potential.
\end{abstract}
\end{frontmatter}


In this letter we present a microscopic calculation of the shear viscosity $\eta$ 
and of its ratio to the entropy density $s$ for a system of quarks. 
This subject has been widely 
discussed in recent times, on the basis of the intriguing results from the 
experiments carried out at the Relativistic Heavy Ion Collider (RHIC). 
In particular the measurement of the $v_2$ coefficient~\cite{exp} in the 
multipole analysis of the angular distribution of the produced hadrons seems 
to imply a very small viscosity, like the one of 
an almost perfect fluid~\cite{Huovinen,Drescher07,Csernai}, in contrast with the 
 current description of the QGP as a gas of weakly interacting 
quasi-particles~\cite{karsch}. The substantial collective flow 
 observed in these collisions also seems to imply 
quite small values for 
the viscosity~\cite{Teaney-shuryak,Teaney-03,Peshier,KapustaQM,Lacey}.

The temperature behavior of the ratio $\eta/s$ should determine the value $T_c$ at 
which a phase transition occurs in the system~\cite{Csernai,Lacey}. Much emphasis 
has received the result of certain special supersymmetric Yang-Mills theories
in 4 dimensions, which predicts a lower limit 
for the viscosity/entropy density ratio, namely $\eta/s\ge 1/4\pi$ (in units 
$\hbar=k_B=c=1$)~\cite{Policastro,Kovtun}; the latter value is approached by several  
recent calculations~\cite{Nakamura,Meyer,Iwasaki,ACHMN}. Transport coefficients
from the linearized Boltzmann equation for a non abelian gauge theory have been
obtained using perturbative QCD in Refs.~\cite{ArnoldI,ArnoldII}.

Here we shall describe the quarks in the QGP within the Hard Thermal Loop (HTL) 
approximation, which has already been proven quite successful for the 
thermodynamics of high temperature QGP~\cite{BIR1,blater,bie}. Quark-antiquark correlators
have also been widely considered in this approach (see, e.g., Refs.~\cite{ABM,ABCM,ABCCM} 
and references therein).

We evaluate the shear viscosity $\eta$ of the system limiting ourselves to the 
quark degrees of freedom and starting from the Kubo formulas for the hydrodynamic 
transport coefficients; it is expressed in terms of the correlator of the 
energy-momentum tensor as follows:
\beq
\eta=\lim_{\omega\to 0^+} \eta(\omega)=
- \left. \frac{\dd}{\dd \omega}
{\mathrm{Im}}\, \Pi^R(\omega) \right|_{\omega=0^+}\,,
\label{eq:visc}
\eeq
where we consider the following retarded correlator (at zero momentum)~\cite{ACHMN}
\beq\label{eq:pi}
\Pi^R(i\omega_l)=N_c N_f\,\,\frac{1}{\beta}\sum_{n=-\infty}^{+\infty}\,\,\int\!\frac{d^3k}{(2\pi)^3}
\,\, k_x^2 \,\, \trm{Tr}[\gamma_{2}\,S(i\omega_n,k)\gamma_{2}\,S(i\omega_n\!-i\omega_l,k)]\;.
\eeq
In the above $\omega_l=2\pi l/\beta$,  $\omega_n=(2n+1)\pi/\beta$, ($\beta=1/T$) 
are Matsubara frequencies and 
the spectral representation of the massless quark propagator is employed
\beq\label{eq:spec}
S(i\omega_{n},k)=-\int\limits_{-\infty}^{+\infty}\!
\frac{d\omega}{2\pi}\,\,\frac{{\rho(\omega,k)}}
{i\omega_n-\omega}\;,
\eeq
 the spectral function being conveniently split into the chirality projections:
\beq\label{spectral}
\rho(\omega,k)=\frac{\gamma^0-\mathbf{{\bm \upgamma}\cdot\hat{k}}}{2}\rho_{+}(\omega,k)\, +\, 
\frac{\gamma^0+\mathbf{{\bm \upgamma}\cdot\hat{k}}}{2}\rho_{-}(\omega,k) \; .
\eeq

The explicit expression of the HTL quark spectral function introduced 
in Eq.(\ref{spectral}) reads~\cite{lb}:
\beq
\rho_{\pm}(\omega,k)=  2 \pi Z_{\pm}(k) [\delta(\omega-\omega_{\pm})
+\delta(\omega+\omega_{\mp})]
+  2 \pi \beta_{\pm}(\omega,k)\theta(k^2-\omega^2)\;,\label{2part}
\eeq
where $m_q=g(T)T/\sqrt{6}$ is the quark thermal mass,
\beq
Z_{\pm}(k)=\frac{\omega_{\pm}^2(k) - k^2}{2 m_q^2}
\eeq
are the residues of the quasi-particle poles [see Eq.~(\ref{QPpoles})] and

\beqa
&&\beta_{\pm}(\omega,k)= -\frac{m_q^2}{2}\times\\
&& \frac{\pm\omega-k}{\displaystyle{\left\{k(-\omega\pm k)+m_q^2
\left[\pm1-\left(\frac{\pm\omega-k}{2k}\right)\ln
\frac{k+\omega}{k-\omega}\right]\right\}^2+
\left\{\frac{\pi}{2} m_q^2\,\left(\frac{\pm\omega-k}{k}\right)\right\}^2}}\;.
\nonumber
\eeqa

The spectral functions $\rho_{\pm}(\omega,k)$ consist of two pieces, a pole term  
and a cut:
\beq
\rho_{\pm}(\omega,k)= \rho_{\pm}^{QP}(\omega,k)+\rho_{\pm}^{LD}(\omega,k)~.
 \label{QPLD}
\eeq
At a given value of the spatial momentum $k$, in the time-like domain ($\omega^2>k^2$) 
discrete poles are associated to quasiparticle (QP) excitations
with dispersion relation $\omega=\omega_\pm(k)$,  while
in the space-like domain ($\omega^2<k^2$) a cut 
accounts for the \emph{Landau damping} (LD)
of a quark propagating in the thermal bath. The two poles 
correspond to quasi-particles with opposite chirality/helicity sign, their dispersion 
relation being given by the solutions of the implicit equation:
\beq
\omega_\pm= \pm p+\frac{m_q^2}{2p}\left[\left(1\mp\frac{\omega_\pm}{p}\right)
\ln\left|\frac{\omega_\pm+p}{\omega_\pm-p}\right|\pm 2\right].
\label{QPpoles}
\eeq
Vertex corrections to the correlator (\ref{eq:pi}) deserve further investigation and
will not be included here. 

We can now insert the spectral representation, Eq.(\ref{eq:spec}), of the propagators 
into Eq.(\ref{eq:pi}) and sum over the Matsubara frequencies with a standard 
contour integration; then, by performing the usual analytic continuation 
$i\omega_{l}\rightarrow\omega +i\eta^+$ (retarded boundary conditions), taking 
the imaginary part of the result and putting into Eq.(\ref{eq:visc}) 
we obtain~\cite{Iwasaki,ACHMN}:
\beq
\eta=\frac{N_cN_f}{2T}\int \frac{d^3 k}{(2\pi)^3}\, k^2_x 
\int_{-\infty}^{+\infty}\frac{d\omega}{2\pi} 
[1-f(\omega)]f(\omega)
\,\tr \left[\gamma^2 \rho(\omega,{\bf k})\gamma^2 \rho(\omega,{\bf k})\right]\,,
\label{eq:etatot}
\eeq
where $f(\omega)=1/(\exp^{\beta(\omega-\mu)}+1)$ is the thermal distribution
for fermions with chemical potential $\mu$.

The trace over spin can be explicitly carried out after inserting Eq.~(\ref{spectral}) 
into Eq.~(\ref{eq:etatot}) and, since
\beq
\trm{Tr}\left[\gamma_2\frac{\gamma^0\mp\mathbf{{\bm \upgamma}\cdot\hat{k}}}{2}\gamma_2
\frac{\gamma^0\mp\mathbf{{\bm \upgamma}\cdot\hat{k}}}{2}\right]= 2 \hat{k}_y^2,\quad
\trm{Tr}\left[\gamma_2\frac{\gamma^0\mp\mathbf{{\bm \upgamma}\cdot\hat{k}}}{2}\gamma_2
\frac{\gamma^0\pm\mathbf{{\bm \upgamma}\cdot\hat{k}}}{2}\right]= 2-2 \hat{k}_y^2 \;,
\eeq
one finally gets:
\beqa
\eta&&=\frac{N_cN_f}{60 \pi^3 T}\int\limits_{0}^{+\infty} d k \, k^4\times
\label{eq:eta}\\
&&\int\limits_{-\infty}^{+\infty}d\omega\,
[1-f(\omega)]f(\omega)
\left\{
\rho_{+}^2(\omega,{\bf k})+ \rho_{-}^2(\omega,{\bf k}) +  
8 \rho_{+}(\omega,{\bf k}) \rho_{-}(\omega,{\bf k})\right\} ~.
\nonumber
\eeqa

By inserting the explicit expressions for $\rho_{\pm}(\omega,k)$ into Eq.~(\ref{eq:eta})
 this quantity turns out to be split, in the HTL approximation, into three different 
contributions [pole-pole term, labelled QP, pole-cut term (QPLD) which mixes quasi-particle 
and Landau damping parts and cut-cut (LD) which comes from Landau damping only]:
\beq
\eta = \eta^{QP} +\eta^{QPLD} + \eta^{LD}~.
\eeq

The quasi-particle (pole-pole) contribution reads:
\beqa
\eta^{QP} &=&\frac{N_cN_f}{60 \pi^3 T}\int\limits_{0}^{+\infty} d k \, k^4 
\int\limits_{-\infty}^{+\infty}d\omega\, [1-f(\omega)]f(\omega)
\left\{\rho_{+}^{QP}(\omega,{\bf k}) - \rho_{-}^{QP}(\omega,{\bf k}) \right\}^2
\nonumber\\
&+&\frac{N_cN_f}{60 \pi^3 T}\int\limits_{0}^{T} d k \, k^4 
\int\limits_{-\infty}^{+\infty}d\omega \, [1-f(\omega)]f(\omega)
\left\{10 \rho_{+}^{QP}(\omega,{\bf k}) \rho_{-}^{QP}(\omega,{\bf k})\right\} ~.
\label{eq:etaQP}
\eeqa
The Landau damping (cut-cut) contribution reads:
\beqa
\eta^{LD} &=&\frac{N_cN_f}{60 \pi^3 T}\int\limits_{0}^{+\infty} d k \, k^4 
\int\limits_{-k}^{+k}d\omega \, [1-f(\omega)]f(\omega) \times \\
&\times& 
\left\{[\rho_{+}^{LD}(\omega,{\bf k})-\rho_{-}^{LD}(\omega,{\bf k})]^2+  
10 \rho_{+}^{LD}(\omega,{\bf k}) \rho_{-}^{LD}(\omega,{\bf k})\right\} \nonumber
\label{eq:etaLD}
\eeqa
and finally the pole-cut contribution  reads:
\beqa
\eta^{QPLD} &=&\frac{N_cN_f}{60 \pi^3 T}\int\limits_{0}^{+\infty} d k \, k^4 
\int\limits_{-k}^{+k} d\omega \, [1-f(\omega)]f(\omega) \times \\
&\times& 
\left\{ 2 \rho_{+}^{QP}(\omega,{\bf k}) \rho_{+}^{LD}(\omega,{\bf k}) +
2 \rho_{-}^{QP}(\omega,{\bf k}) \rho_{-}^{LD}(\omega,{\bf k}) \right.  \nonumber \\
&+&
\left. 8 \rho_{+}^{QP}(\omega,{\bf k}) \rho_{-}^{LD}(\omega,{\bf k}) +
8 \rho_{-}^{QP}(\omega,{\bf k}) \rho_{+}^{LD}(\omega,{\bf k})
\right\}~. \nonumber
\label{eq:etaQPLD}
\eeqa

Among the three contributions, the QP one, as it stands, turns out to be
divergent, as it corresponds to a free gas of particles. 
In order to cure this unphysical outcome, we shall assign,
following Ref.~\cite{BP}, a finite width to the poles keeping unchanged 
the residues: the width of the poles $\pm\omega_+(k)$
is $\gamma_+$ and the width of $\pm\omega_-(k)$ is $\gamma_-$. The spectral
function for these dressed quasiparticles then reads:
\beq\label{eq:rhoDQP}
\rho_{\pm}^{\textrm{QP}}
(\omega,{\bf k})= Z_{\pm}(k)
\frac{2\gamma_{\pm}}{[\omega-\omega_\pm(k)]^2+\gamma_\pm^2}
+Z_{\mp}(k)
\frac{2\gamma_{\mp}}{[\omega+\omega_\mp(k)]^2+\gamma_\mp^2}
\eeq
and is normalized according to
\beq\label{eq:norm}
\int\limits_{-\infty}^{+\infty} \frac{d\omega}{2\pi}
\rho_{\pm}^{\textrm{QP}}
(\omega,{\bf k})=Z_+(k)+Z_-(k)~.
\eeq
The width is fixed as in Ref.~\cite{BP}:
\beq
\gamma_{\pm} = a \frac{C_f g^2 T}{16 \pi} ,
\eeq
\
where $a= 3.0$ is chosen as an intermediate value
between the one loop calculation and the improved
one of Ref.~\cite{BP} for $N_f =2$. $C_f=4/3$ is the Casimir constant for the
fundamental representation of SU($N_f$).\footnote{An additional reason for the present choice 
lies in the fact that in Ref.~\cite{BP} the width is calculated at zero momentum, 
while we are integrating on the quasi-particle contribution over all momenta.}
A word of caution is in order with the use of a constant quasi-particle width:
indeed, as observed in Ref.~\cite{ACHMN}, the second term of $\eta^{QP}$
[second line in Eq.~(\ref{eq:etaQP})] would contain an ultraviolet
divergence. A natural cutoff is provided by the scale of momenta within the
HTL approach, since the particles dressed by HTL self-energy are expected to be
``soft'' constituents, where the soft momenta are defined as the ones smaller 
than $T$ (or $gT$). Hence the upper limit of the integration over  $k$ in the 
second line of Eq.~(\ref{eq:etaQP}) has been set to $T$.

In addition to the shear viscosity, we have then evaluated  the entropy density  
according to the formulas of Ref.~\cite{blater,bie}. 

\begin{figure}[h]    
\begin{center}
{
\includegraphics[clip,width=0.7\textwidth]{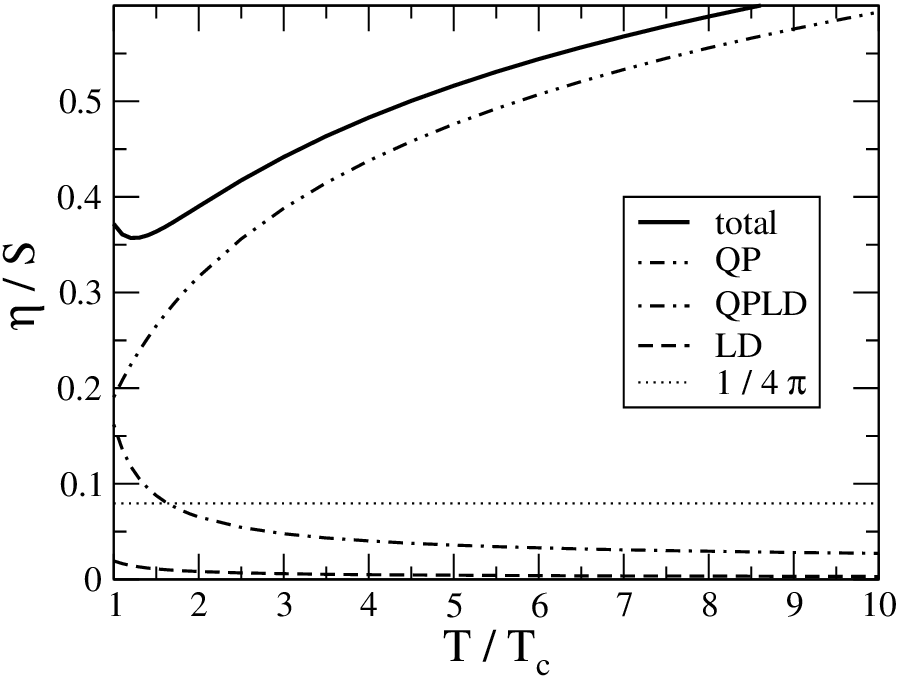}    
}
\end{center}
\caption{The ratio $\eta/s$ for $\mu=0$ as a function of $T/T_c$: the QP (dash-double-dotted line), 
LD (dashed) and the interference between QP and LD (dot-dashed) are shown, together with
the total result (solid). The limit $1/4\pi$ is indicated by the dotted line.}
\label{fig:01}   
\end{figure}

The ratio $\eta/s$ is illustrated in Fig.~\ref{fig:01} (solid line) for $\mu=0$ 
as a function of $T/T_c$, 
with $T_c=202$~MeV, which was suggested as the transition temperature for $N_f=2$ 
in Ref.~\cite{kacz}. Moreover we use the temperature dependent running gauge coupling 
given by the two-loop perturbative beta-function~\cite{karsch,goc,zan,ABCM}. 
In the same figure we display the separate contributions from $\eta^{QP}$, 
$\eta^{QPLD}$ and $\eta^{LD}$. 
Clearly the dominant contribution is $\eta^{QP}$: we notice that the cutoff 
dependence in the second term of the QP contribution is quite mild, since the
first term is by far dominant; indeed by artificially increasing the cutoff up
to $10 T$ the ratio $\eta/s$ increases by about $2\%$.
We wish to stress that the contribution of the continuum
 to the viscosity was never evaluated in previous works: we find that
it is quite small and that its temperature dependence is opposite to the QP one, showing a 
fast decrease, which is also reflected in the QP-LD interference term. 

\begin{figure}[h]   
\begin{center}
{
\includegraphics[clip,width=0.7\textwidth]{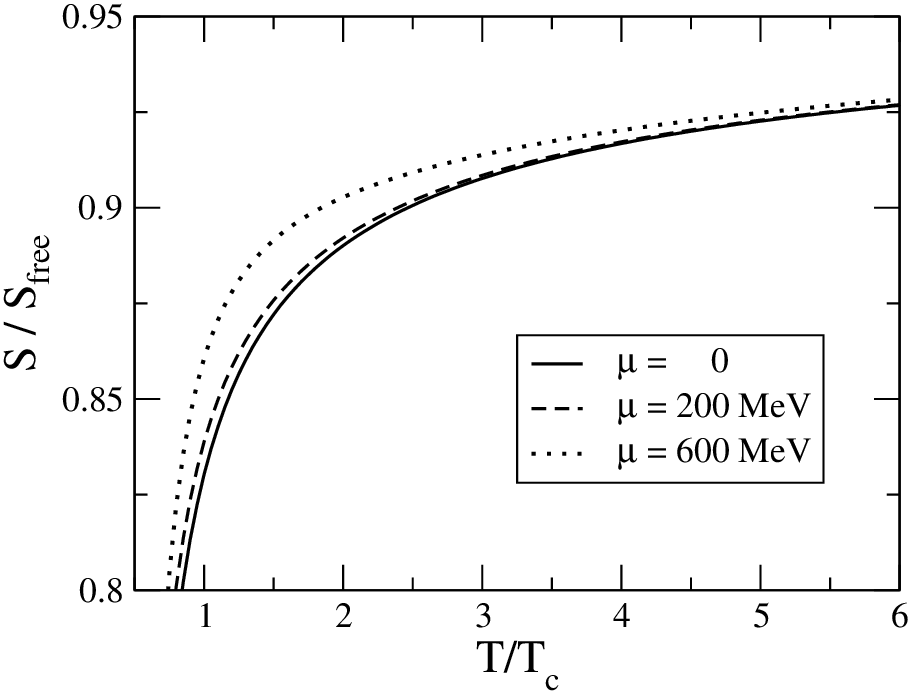}    
}
\end{center}
\vspace{-0.4cm}
\caption{Entropy density of quarks in HTL approximation (divided by $s_{\mathrm{free}}$)
  as a function of $T/T_c$ for  $\mu=0$ (solid line), $\mu=200$~MeV 
(dashed) and  $\mu=400$~MeV (dotted).}   
\label{fig:02}   
\end{figure}

We have then extended our calculation of $s$, $\eta$ and $\eta/s$ to the 
case of finite chemical potential, taking also into account the corresponding 
correction to the quark mass~\cite{BB}, ~
$m_q^2={g^2(T)}\left(T^2+{\mu^2}/{\pi^2}\right)/{6}$. 
Finite baryonic densities in high energy heavy ion collisions will be of interest
for the forthcoming GSI experiments. The ratio of the 
entropy density to the corresponding Stefan-Boltzmann limit ($s_{\mathrm{free}}$) is shown in 
Fig.~\ref{fig:02} for three different values of $\mu$: the curves
corresponding to $\mu\ne 0$ have been extended to lower temperatures, since 
$T_c$ should gradually decrease with increasing chemical potential. 
The effect of the chemical potential appears to be sizable only for the
largest value considered here.

\begin{figure}[h]    
\begin{center}
{
\includegraphics[clip,width=0.7\textwidth]{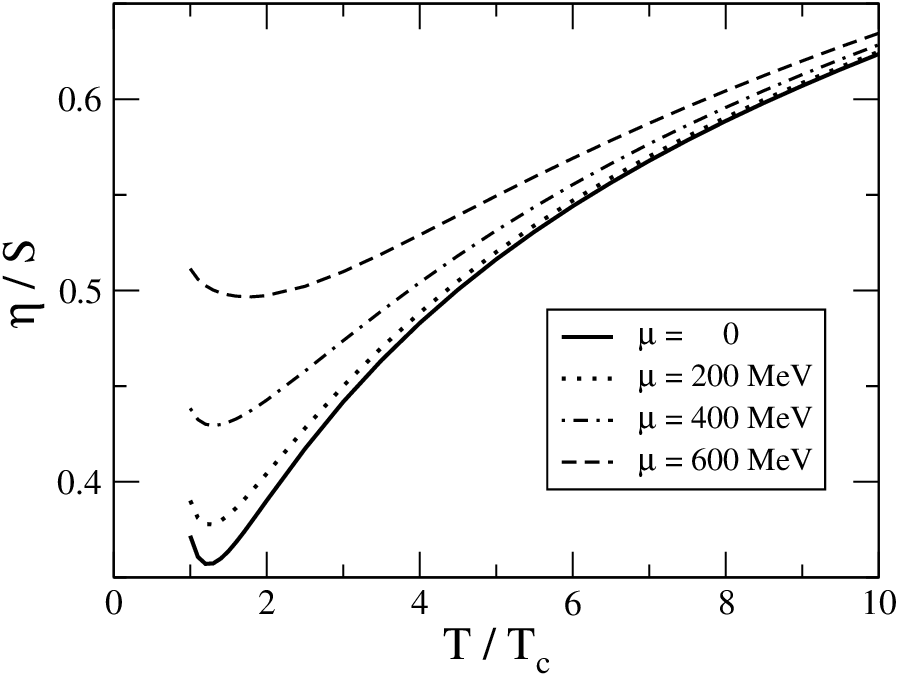}    
}
\end{center}
\caption{The ratio $\eta/s$ as a function of $T/T_c$ for different values of the chemical 
potential: $\mu=0$ (solid line), $\mu=200$~MeV (dotted),  $\mu=400$~MeV (dot-dashed)
and $\mu=600$~MeV (dashed).}
\label{fig:03}   
\end{figure}

In Fig.~\ref{fig:03} we display the $\eta/s$ ratio  for different 
values of $\mu$: one can see that the effect of a finite chemical potential increases 
the ratio, which implies a larger correction to the viscosity than to the entropy 
density (the modification of the latter, according to the results shown in Fig.~\ref{fig:02}, 
should lower $\eta/s$). Obviously the largest effects are seen at lower temperatures.

%
In summary we have microscopically calculated the viscosity and the entropy 
density for a system of quarks described in the framework of the HTL
approximation. The quasi-particle term in the spectral density has been 
modified by accounting, within the same approach, for a finite width, which
removes the divergence in the shear viscosity derived from the Kubo formula.
The results for the $\eta/s$ ratio are comparable to the ones obtained within
phenomenological models~\cite{ACHMN} and turn out to be about a factor 3 larger 
than recent lattice calculations~\cite{Meyer} in the pure gauge sector.
We found that the continuum part of the quark spectral
density gives a negligible contribution to the viscosity and the corresponding 
$\eta/s$ ratio slowly decreases with temperature, at variance with the dominant, 
quasi-particle one.

We have also considered the effect of a finite chemical potential and we found 
that it causes a sizable increase, both in the entropy density and in the viscosity: 
indeed $\mu$ induces the largest corrections to the latter, so that also the ratio $\eta/s$ 
tends to increase with the chemical potential. 
\vspace{1cm}

{\bf Acknowledgments}
One of the authors (P.C.) thanks the Department of Theoretical Physics of the Torino University 
and INFN, Sezione di Torino, for the warm hospitality.

\end{document}